\documentclass{iopart}
\usepackage{graphicx}
\usepackage{iopams}
\usepackage{harvard}
\begin{document}

\title{Quantum teleportation and computation with Rydberg atoms in an optical lattice}

\author{Huaizhi Wu, Zhen-Biao Yang, Li-Tuo Shen, and Shi-Biao Zheng}

\address{Department of Physics, Fuzhou University, Fuzhou 350002, People's Republic of China}

\begin{abstract}
Neutral atoms excited to Rydberg states can interact with each other via dipole-dipole interaction, which results in a physical phenomenon named Rydberg blockade mechanism. The effect attracts much attention due to its potential applications in quantum computation and quantum simulation. Quantum teleportation has been the core protocol in quantum information science playing a key role in efficient long-distance quantum communication. Here, we first propose the implementation of teleportation scheme with neutral atoms via Rydberg blockade, in which the entangled states of qubits can readily be prepared and the Bell states measurements just require single qubit operations without precise control of Rydberg interaction. The rapid experimental progress of coherent control of Rydberg excitation, optical trapping techniques and state-selective atomic detection promise the application of the teleportation scheme for scalable quantum computation and many-body quantum simulation using the protocol proposed by D. Gottesman and I. L. Chuang [Nature (London) 402, 390 (1999) ] with Rydberg atoms in optical lattice.

\end{abstract}

\pacs{03.67.Lx, 03.67.Ac, 32.80.Ee, 42.50.Dv}
%03.67.Lx	Quantum computation architectures and implementations
%42.50.Dv	Quantum state engineering and measurement
%03.67.Ac	Quantum algorithms, protocols, and simulations
%32.80.Ee	Rydberg states
\submitto{\JPB}
\maketitle

Quantum teleportation, arises from non-local property of quantum mechanics, allows for efficient transmission of quantum information from one location to another with prior sharing of an Einstein-Podolsky-Rosen (EPR) pair and a conventional communication channel \cite{Bennett_PRL1993}. Teleportation is expected to play an important role in quantum communication \cite{Duan_Nature2001} and quantum computation \cite{Gottesman_Nature1999}. Using quantum entanglement as a resource, teleportation can be used for building quantum logic modules for universal quantum computation. Although manufacture of a practical quantum computer is out of reach so far, simulation of many-body interaction (e.g., Ising model) that is of currently great interest can be done with quantum teleportation as well \cite{Dur_PRA2008}. The teleportation-based method together with entanglement purification can simulate interacting high dimensional quantum systems and reduce the influence of quantum noise \cite{Dur_PRA2008}. Experimental realization of quantum entangling gates based on quantum teleportation has been demonstrated with linear optics system \cite{Pan_PNAS2011}. 

Neutral atoms trapped in optical lattice provides an architecture for effective quantum control \cite{Bloch_RMP2008,Bloch_NatPhy2012}. Neutral atom excited to high-lying electronic states is referred to as Rydberg atom \cite{Gallagher_Rydberg_Atom}. Atoms in Rydberg states have the size $r\sim n^2 a_0$ with $n$ the principle quantum number and $a_0$ the Bohr radius, which gives rise to large electric dipole moments and thus great sensitivity to external electric fields. The radiative lifetime $\tau$ of the Rydberg states has a scaling of $\tau\sim n^{3}$ for low angular momentum states (typically tens of $\mu$s) and is influenced by surrounding temperature. Rydberg-Rydberg interaction arises when Rydberg atoms are exposed to external applied fields and the distance between Rydberg atoms is within a characteristic interaction range $r_{0}$. For strong Rydberg-Rydberg interaction, excitation of an atom to Rydberg state using a laser light can inhibit the other within $r_{0}$ from excitation by introducing an energy shift for the double-excitation state, which is known as Rydberg blockade mechanism \cite{Vogt_PRL2007,Jaksch_PRL2000,Lukin_PRL2001}. An alternative mechanism causing blockade effect is resonant F\"orster interaction \cite{Forster_Physik1948,Urban_NatPhy2009}. These effects have been widely explored for their potential applications in quantum information processing \cite{Jaksch_PRL2000,Lukin_PRL2001,Saffman_RMP2010,Comparat_JOSAB2010,Moller_PRL2008,Muller_PRL2009,Brion_PRA2012,Wu_PRA2010,Zhao_PRA2010_repeater,Brion_PRL2008_ErrorC,Han_PRA2010_repeater}, quantum simulation of the interacting Rydberg gas \cite{Weimer_NatPhy2010,Lee_PRA2011,Ji_PRL2011,Qian_PRA2012} and studying interaction-induced optical nonlinearity \cite{Gorshkov_PRL2011,Petrosyan_PRL2011,Parigi_PRL2012,Pritchard_PRL2012}. Experimental demonstrations for preparation of entangled states and implementation of quantum logic gates has been reported\cite{Isenhower_PRL2010,Wilk_PRL2010}. Blockade effect is found to be significant as well for other quantum systems, such as electron spins \cite{Koppens_Nature2006,Shaji_NatPhy2008} and cold polar molecules \cite{JinYe_PhysicsToday2011}.

\begin{figure}
\centering
\includegraphics[width=0.8\columnwidth]{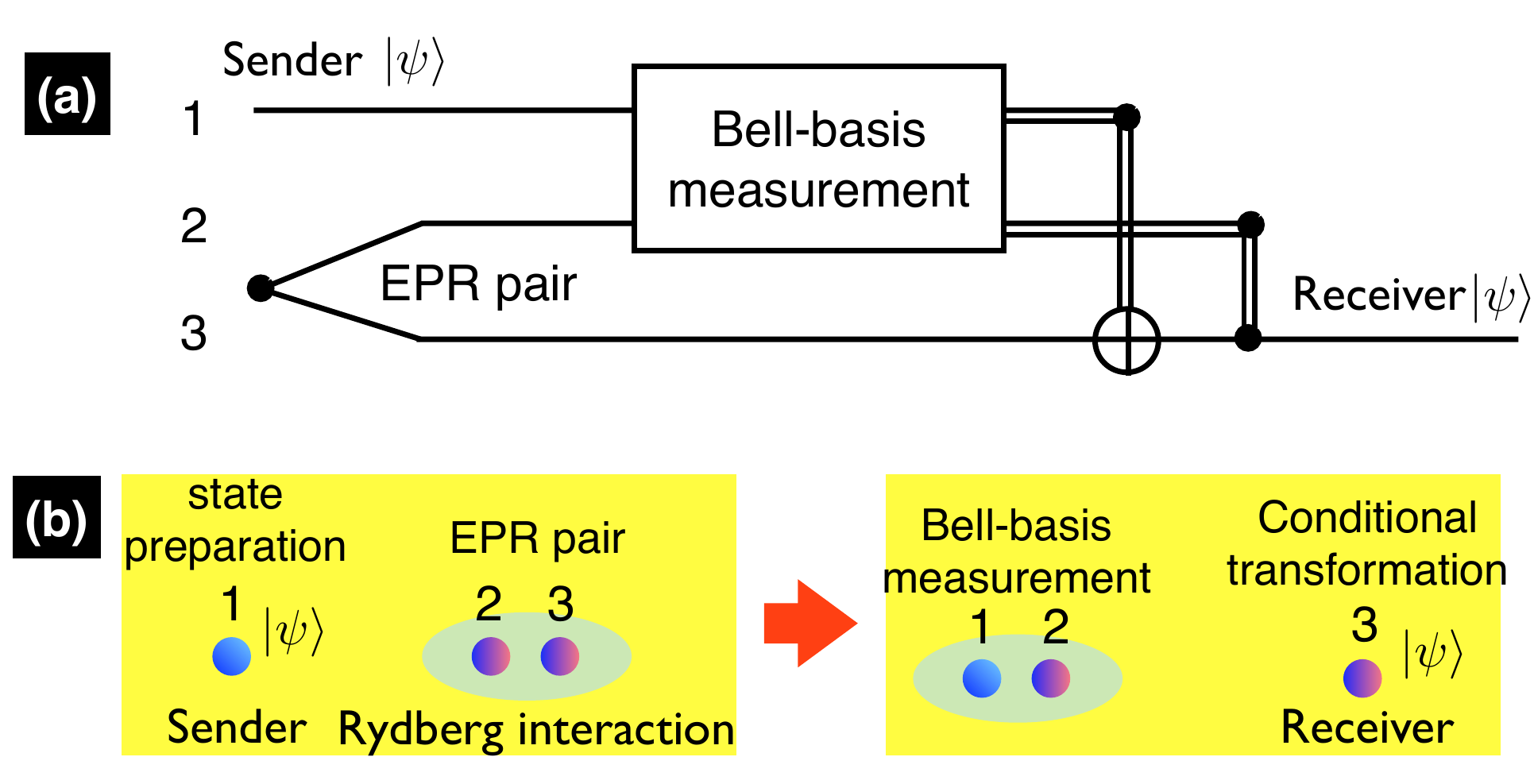}
%\Figure{(a) Quantum circuit representation of quantum teleportation. (b) Physical realization of quantum teleportation with neutral Rydberg atoms.}
\caption{\label{fig:Fig1_Quantum_Circuit} (a) Quantum circuit representation of quantum teleportation. (b) Physical realization of quantum teleportation with neutral Rydberg atoms.}

\end{figure}

In this paper, we first propose a scheme for implementing quantum teleportation with neutral atom qubits, which interact with each other via strong and long-range dipole-dipole or van der Waals interaction. The merits of our protocol include (1) easy preparation of EPR states, (2) quantum logic operations without precise control of Rydberg-Rydberg interaction, (3) Bell states measurements involving only single qubit operations and state-selective detection based on conditional state transfer. We then outline a protocol where teleportation can be implemented in Rydberg atom array trapped in periodic optical potential to complete teleportation-based quantum computation and quantum simulation. The experimental demonstration of single-site-resolved optical control in an optical lattice paves the way for the implementation of our protocol \cite{Bloch_NatPhy2012}.

The quantum circuit representation of teleportation is shown in Fig. \ref {fig:Fig1_Quantum_Circuit}(a) \cite{Bennett_PRL1993}. There are three qubits involved. The quantum information to be teleported is carried by qubit 1. The qubits 2 and 3 are initially prepared in an entangled EPR state. Performing a Bell measurement on qubits 1 and 2 by the sender yields two classical bits of information, according to the outcomes, the receiver can apply a suitable single-qubit operation on qubit 3 to reconstruct the initial state of qubit 1 and to obtain the desired identical quantum information given by qubit 1. In our scheme, neutral atoms (for example, $^{87}$Rb), which make long range interaction while they are excited to the Rydberg state, are employed as quantum bits (see Fig. \ref{fig:Fig1_Quantum_Circuit}(b)).

\begin{figure}
\centering
\includegraphics[width=0.8\columnwidth]{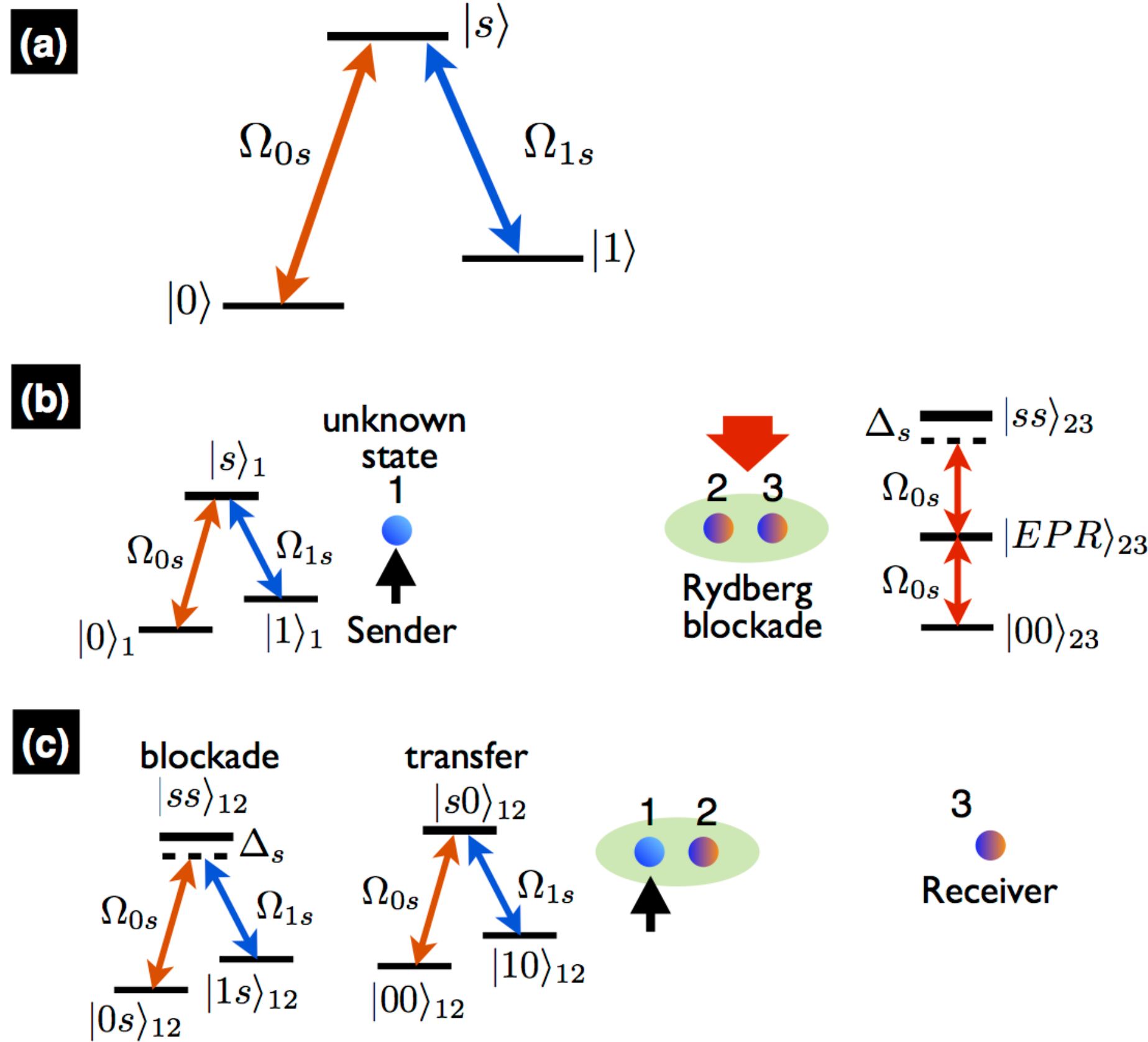}
\caption{\label{fig:Fig2_EnergyLevels}(a) Schematic energy level structure of the atomic qubits. Two ground states $|0\rangle$ and $|1\rangle$ couple to the high-lying Rydberg state $|r\rangle$ with Rabi frequencies $\Omega_{0r}$ and $\Omega_{1r}$, respectively. (b) State preparation of the teleported qubit and the laser manipulation scheme for the EPR pair preparation. Effective two-atom energy levels including Rydberg blockade is shown. (c) Conditional state transfer and blockade of the qubit 1.}
\end{figure}

Neutral atoms are trapped in optical lattices or optical tweezers and are frozen to the motional ground state. The structure of the relevant energy levels of the atoms is schematically shown in Fig. \ref{fig:Fig2_EnergyLevels}(a). Two ground states denoted by $|0\rangle$ and $|1\rangle$ are coupled to the Rydberg excited state $|r\rangle$ by two lasers with Rabi frequencies $\Omega_{0r}$ and $\Omega_{1r}$, respectively. While two neutral atoms are exposed to the common laser beams used for excitation of the Rydberg state $|r\rangle$, the two-atom Rydberg-Rydberg interaction gives rise to an energy shift denoted by $\Delta_{r}$, which is determined by the principle quantum number of the Rydberg state and the interatomic distance. In experiment, the Rydberg excitation can be realized by two-photon transitions or by short wavelength single photon transition.

\begin{figure}
\includegraphics[width=0.8\columnwidth]{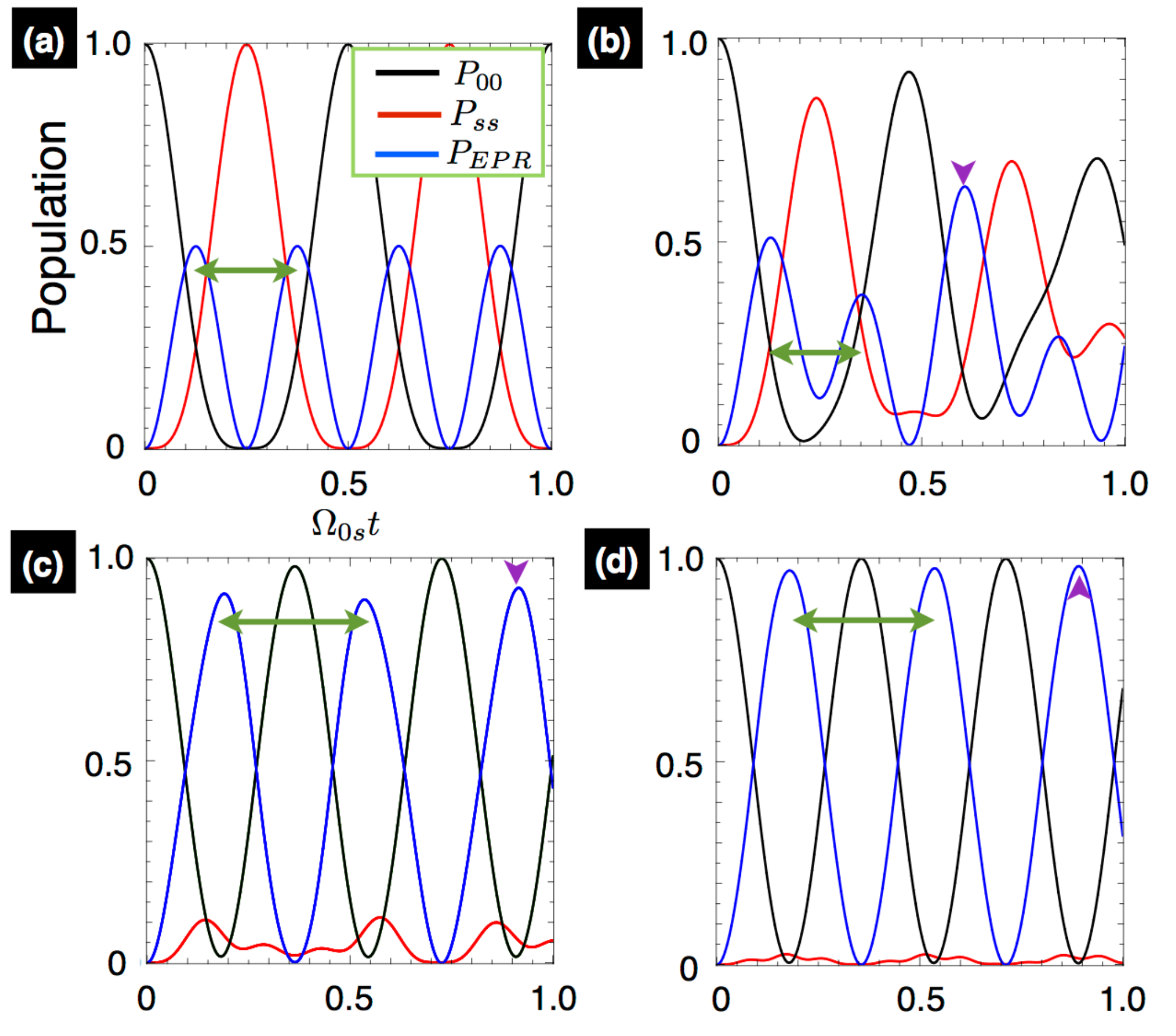}
\centering
\caption{\label{fig:Fig3_Fepr_1d} Population of atomic states $|00\rangle_{23}$, $|EPR\rangle_{23}$, and $|rr\rangle_{23}$ versus dimensionless rescaled time $\Omega_{0r}t$ with Rydberg interaction strength (a) $\Delta_{r}/\Omega_{0r}=0$, (b)$\Delta_{r}/\Omega_{0r}=1$, (c)$\Delta_{r}/\Omega_{0r}=5$, and (d)$\Delta_{r}/\Omega_{0r}=10$. The initial state is $|00\rangle_{23}$. The effect of spontaneous emission from Rydberg state is neglected here. The Rydberg interaction strength varies the time evolution of the two-atom system. Without interaction the system displays a perfect sinusoidal oscillation, see (a). The gradually enhanced interaction strength hinders the atoms from double excitation. The population $P_{EPR}$ of the maximally entangled state $|EPR\rangle_{23}$ can exceed 0.95 for $\Delta_{r}/\Omega_{0r}=10$. As $\Omega_{0r}t$ grows, the peak value of $P_{EPR}$
  (indicated by purple arrows) can be higher than other peaks without considering spontaneous emission.}
\end{figure}

\begin{figure}
\centering
\includegraphics[width=0.8\columnwidth]{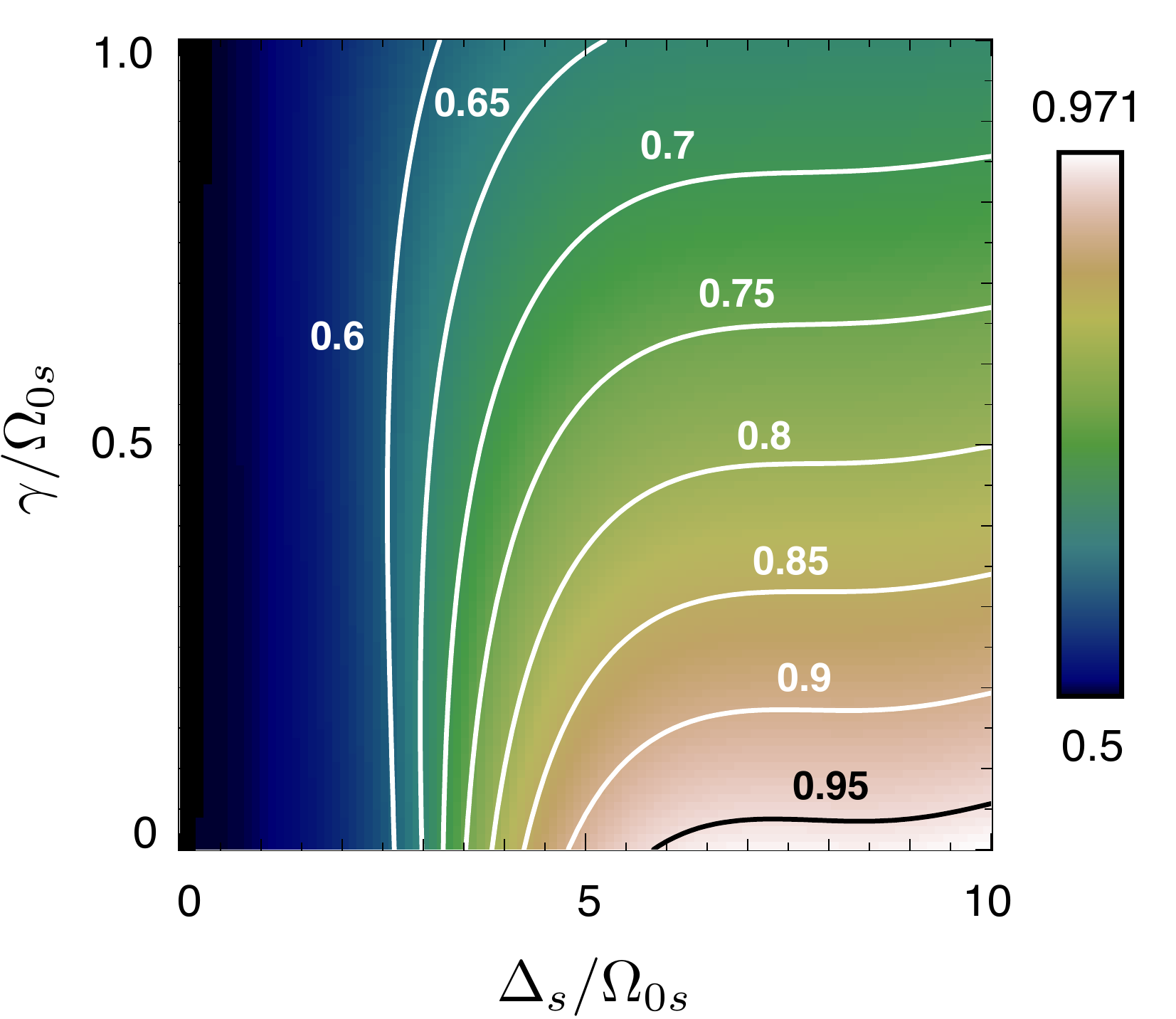}
\caption{\label{fig:Fig4_Fepr_2d} Probability for obtaining two-atom maximally entangled states $|EPR\rangle_{23}$ versus Rydberg interaction strength $\Delta_{r}/\Omega_{0r}$ and spontaneous decay $\gamma/\Omega_{0r}$ of the Rydberg state $|r\rangle$. The spontaneous emission rate $\gamma$ and the Rydberg interaction strength $\Delta_{r}$ are both related to the principle quantum number of the Rydberg state $|r\rangle$. The elaborate selection of Rabi frequency $\Omega_{0r}$ and excited energy level $|r\rangle$ leads to high fidelity preparation of $|EPR\rangle_{23}$.}
\end{figure}

Suppose the system consisting of three atomic qubits is initially in the state $(\alpha|0\rangle_{1}+\beta|1\rangle_{1})|0\rangle_{2}|0\rangle_{3}$. The qubit 1 has been prepared in an unknown quantum state (see left-hand side, Fig. \ref{fig:Fig2_EnergyLevels}(b)). The teleportation scheme in general consists of three procedures. (1) Entanglement preparation. So far, entanglement of individual neutral atoms via Rydberg blockade can be experimentally realized in two different ways. One is deterministic generation of entangled states using Rydberg blockade mediated controlled-NOT gate \cite{Isenhower_PRL2010,Zhang_PRA10_esCNOT}. The other is direct generation of EPR pair depending on Rydberg-excitation competition between two ground-state neutral atoms \cite{Wilk_PRL2010}. Here, we assume entanglement of atom pair is prepared by the latter method. The qubits 2 and 3 within Rydberg interaction range are manipulated by a common laser pulses sequence (sketch map shown in right-hand side, Fig. \ref{fig:Fig2_EnergyLevels}(b)). The time evolution of the system, in the interaction picture, can be described by the Hamiltonian
 
 \begin{equation}
 H_{23}=\sum_{j=2,3}(\Omega_{0r}|r\rangle_{jj}\langle0|+\Omega_{0r}^{*}|0\rangle_{jj}\langle r|)+\Delta_{r}|r\rangle_{2}|r\rangle_{33}\langle r|_{2}\langle r|,
 \end{equation} where $j=2, 3$, and we assume the interaction of the atoms with laser fields are identical so that they have the identical Rabi frequency $\Omega_{0r}$, which suppose to be real for simplification later. For $\Delta_{r}\gg\Omega_{0r}$, only one of the qubits will be excited to the Rydberg state with resonant pulse length, the strong Rydberg blockade effect prevents the other qubit from excitation, namely, it remains in the ground state $|0\rangle$. The effective Hamiltonian for the process is given by
 $H_{23}^{eff}=\sqrt{2}\Omega_{0r}[\frac{1}{\sqrt{2}}(|0\rangle_{2}|r\rangle_{3}+|r\rangle_{2}|0\rangle_{3})_{2}\langle0|_{3}\langle0|]+h.c.$. Thus, choosing appropriate interaction time the system dominated by $H_{23}^{eff}$ evolves into
 
  \begin{equation}
 (\alpha|0\rangle_{1}+\beta|1\rangle_{1})\otimes\frac{1}{\sqrt{2}}(|0\rangle_{2}|r\rangle_{3}+|r\rangle_{2}|0\rangle_{3}). \label{eq:system_step1}
  \end{equation} The fidelity of the entangled state $|EPR\rangle_{23}=(|0\rangle_{2}|r\rangle_{3}+|r\rangle_{2}|0\rangle_{3})/\sqrt{2}$ is essentially limited by the interaction strength $\Delta_{r}$. In the dispersive regime, the transition from $|EPR\rangle_{23}$ to $|rr\rangle_{23}$ can be basically inhibited. We show the population of atomic states $|00\rangle_{23}$, $|EPR\rangle_{23}$, and $|rr\rangle_{23}$ as a function of rescaled time $\Omega_{0r}t$ in Fig.\ref{fig:Fig3_Fepr_1d}. 
  The time-dependent population $P_{EPR}$ of the entangled state $|EPR\rangle_{23}$ oscillates with varied time period revised by $\Delta_{r}$. $P_{EPR}$ increases as Rydberg interaction $\Delta_{r}$ goes stronger. The increasing energy shift $\Delta_{r}$ lowers the probability for detecting double excitation state $|rr\rangle_{23}$ correspondingly, which can be suppressed to 0.05 when the blockade strength reaches $\Delta_{r}/\Omega_{0r}=10$. In addition, we note that the third peak value of $P_{EPR}$ indicated by purple arrows in each sub-figure is larger than other extrema. It means that we can prepare EPR pair with higher fidelity in a longer time limit. However, this would not be the case if the spontaneous decay from excited state $|r\rangle$ is taken into account. Study of the open system including spontaneous emission can be done by using the master equation with the Lindblad form 
 \begin{equation}
 \dot{\rho}_{23}=\frac{1}{i\hbar}[H_{23},\rho_{23}]+\frac{\gamma}{2}\sum_{j=2,3}(2S^{(0)}_{j}\rho_{23}S_{j}^{(0) \dagger}-\rho_{23}S_{j}^{(0) \dagger}S^{(0)}_{j}-S_{j}^{(0) \dagger}S^{(0)}_{j} \rho_{23}),\end{equation} where $S^{(0)}_{j}=|0\rangle_{jj}\langle r|$, $\rho_{23}$ is the system's density matrix, and $\gamma$ is the rate of spontaneous emission. The probability for preparing two-atom maximally entangled state $|EPR\rangle_{23}$ as functions of Rydberg interaction strength $\Delta_{r}$ and spontaneous emission rate $\gamma$ is shown in Fig.\ref{fig:Fig4_Fepr_2d}. Recession of the fidelity of $|EPR\rangle_{23}$ is due to the spontaneously atomic transition to the ground state $|0\rangle$ followed by a photon emitted at random directions as well as Rydberg interaction induced double excitation. Thus, careful selection of atom-field interaction (Rabi frequency) and excited Rydberg energy levels can prompt a high fidelity preparation of entangled states. Besides, the trapping potential for ground and Rydberg states may be different regardless of the type of trap used to hold the atoms, which will lead to motional heating. The effect of the decoherence due to motional heating on the fidelity can be minimized by carefully selecting the frequency of the lattice, so that the ground states and the Rydberg state have the same polarizability, which has been studied by  Safronova et al. \cite{Safronova_PRA2003}.

In terms of the Bell basis states for qubits 1 and 2 in the subspace spanned by $\left\{  |0\rangle_{1}|0\rangle_{2},|0\rangle_{1}|r\rangle_{2},|1\rangle_{1}|0\rangle_{2},|1\rangle_{1}|r\rangle_{2}\right\}$  \begin{equation}
|\Phi_{\pm}^{1,2}\rangle=\frac{1}{\sqrt{2}}(|0\rangle_{1}|0\rangle_{2}\pm|1\rangle_{1}|r\rangle_{2})
  \end{equation}
 and \begin{equation}
 |\Psi_{\pm}^{1,2}\rangle=\frac{1}{\sqrt{2}}(|0\rangle_{1}|r\rangle_{2}\pm|1\rangle_{1}|0\rangle_{2}),
   \end{equation} the system state in Eq.~(\ref{eq:system_step1}) can be rewritten as		
 \begin{eqnarray}
 \frac{1}{2}[|\Phi_{+}^{1,2}\rangle(\alpha|r\rangle_{3}+\beta|0\rangle_{3})+|\Phi_{-}^{1,2}\rangle(\alpha|r\rangle_{3}-\beta|0\rangle_{3}) \nonumber \\
	+|\Psi_{+}^{1,2}\rangle(\alpha|0\rangle_{3}+\beta|r\rangle_{3})+|\Psi_{-}^{1,2}\rangle(\alpha|0\rangle_{3}-\beta|r\rangle_{3})].
		  \end{eqnarray} The measurement of Bell states will collapse the qubit 3 to the quantum states that contain quantum information from qubit 1.

(2) Bell state measurement. Based on the Rydberg blockade mechanism, conditional state transfer and blockade between ground states $|0\rangle$ and $|1\rangle$ for qubit 1 can be realized by addressing the qubit individually, which is utilized for disentangling the Bell basis states. The qubits 1 and 2 are separately measured eventually. In this step, we first switch on the coupling of the Rydberg state  $|r\rangle_{1}$ to both ground states $|0\rangle_{1}$ and $|1\rangle_{1}$. The atom-field interaction is described by the Hamiltonian 
\begin{equation}
 H_{12}=(\Omega_{0r}|r\rangle_{11}\langle0|+\Omega_{1r}|r\rangle_{11}\langle1|+h.c.)+\Delta_{r}|r\rangle_{1}|r\rangle_{22}\langle r|_{1}\langle r|, 
  \end{equation} where $\Omega_{kr}$ ($k=0,1$) are Rabi frequencies with respect to the interaction of laser fields with atomic transitions $|k\rangle_{j}\leftrightarrow|r\rangle_{j}$. Atoms in the excited Rydberg state $|r\rangle$ will spontaneously transit to the ground states via two independent channels, namely, $|r\rangle\rightarrow|0\rangle$ and $|r\rangle\rightarrow|1\rangle$. The corresponding spontaneous emission rates are $\gamma_{0r}$ and $\gamma_{1r}$, respectively. For simplicity we assume $\Omega_{0r}=\Omega_{1r}=\Omega_{01}$ and $\gamma_{0r}=\gamma_{1r}=\gamma$ in the following.
  
 \begin{figure}
 \centering
\includegraphics[width=0.8\columnwidth]{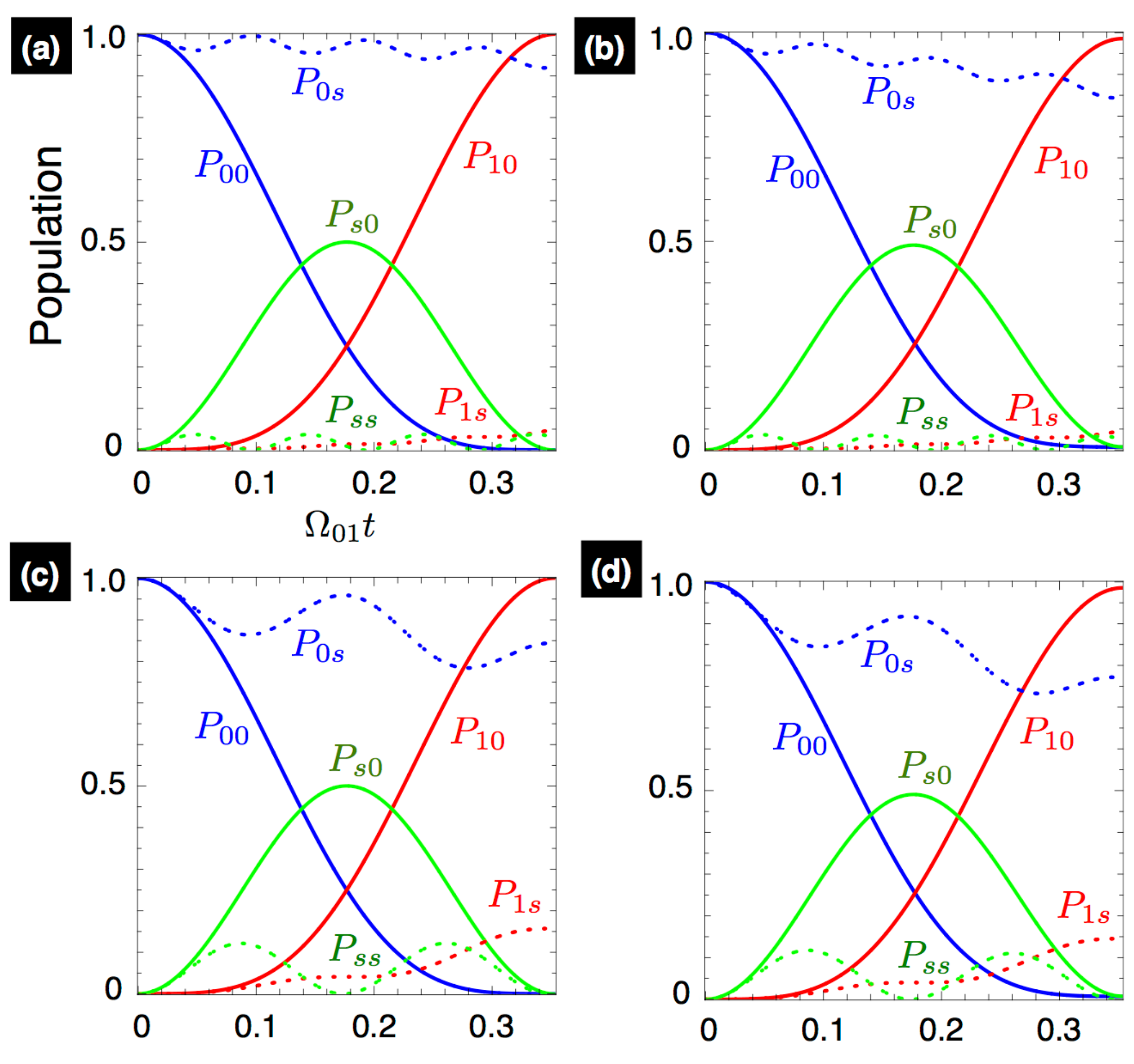}
\caption{\label{fig:Fig5} Disentanglement of qubits 1 and 2 based on conditional state transfer and blockade. Parameters are (a) $\gamma_{01}/\Omega_{01}=0$, $\Delta_{r}/\Omega_{01}=10$, (b) $\gamma_{01}/\Omega_{01}=0.02$, $\Delta_{r}/\Omega_{01}=10$, (c) $\gamma_{01}/\Omega_{01}=0$, $\Delta_{r}/\Omega_{01}=5$, (d) $\gamma_{01}/\Omega_{01}=0.02$, $\Delta_{r}/\Omega_{01}=5$. Without dissipation, transferring from initial state $|0\rangle_{1}|0\rangle_{2}$ to $|1\rangle_{1}|0\rangle_{2}$ is perfect (solid). The system in the state $|0\rangle_{1}|r\rangle_{2}$ is blocked depending on the Rydberg interaction strength $\Delta_{r}$ (dash). Involving spontaneous emission reduces the fidelity of the disentangled operation.}
\end{figure}
  
 The time-evolutional dynamics of the system occurs in two different ways conditionally depending on the excitation status of the qubit 2. Single qubit operations act only on the qubit 1. First, while the atom 2 stays in the Rydberg state $|r\rangle_{2}$, the excitation of qubit 1 to $|r\rangle_{1}$ is impossible if the Rydberg-Rydberg interaction is strong enough. Thus, ideally, both the system states $|0\rangle_{1}|r\rangle_{2}$ and $|1\rangle_{1}|r\rangle_{2}$ will be kept unperturbed. However, due to the finite energy gap $\Delta_{r}$, the transitions out of these two states cannot be completely suppressed. To see this, we find the analytical solution for the system's time evolution with the initial state $|0\rangle_{1}|r\rangle_{2}$\begin{eqnarray}
  & &	\frac{1}{2}[1+e^{-i\frac{\Delta_{r}t}{2}}(\cos(\frac{\Omega^{^{\prime}}t}{2})+i\frac{\Delta_{r}}{\Omega^{^{\prime}}}\sin(\frac{\Omega^{^{\prime}}t}{2}))]|0\rangle_{1}|r\rangle_{2} \nonumber \\
		&+&\frac{1}{2}[e^{-i\frac{\Delta_{r}t}{2}}(\cos(\frac{\Omega^{^{\prime}}t
}{2})+i\frac{\Delta_{r}}{\Omega^{^{\prime}}}\sin(\frac{\Omega^{^{\prime}}t}{2}
))-1]|1\rangle_{1}|r\rangle_{2} \nonumber\\
		&-&ie^{i\frac{\Delta_{r}t}{2}}\frac{\Omega_{01}}{\Omega^{^{\prime}}}\sin(\frac{
\Omega^{^{\prime}}t}{2})|r\rangle_{1}|r\rangle_{2}, \label{eq:q12_CSTransfer}, \end{eqnarray}
where $\Omega^{^{\prime}}=\sqrt{2\Omega_{01}^{2}+\Delta_{r}^{2}}$. For $\Delta_{r}\gg\Omega_{01}$, the state evolution above approximates to $|0\rangle_{1}|r\rangle_{2}\rightarrow\cos(\Omega_{01}^{2}t/4\Delta_{r})|0\rangle_{1}|r\rangle_{2}-i\sin(\Omega_{01}^{2}t/4\Delta_{r})|1\rangle_{1}|r\rangle_{2}$,  from which we can derive the probability for the system staying in the initial state $(1+\cos(\Omega_{01}^{2}t/2\Delta_{r}))/2$. Exact results can be alternatively obtained via numerical simulation, as shown in Fig.\ref{fig:Fig5}. The blockade effect strongly depends on the strength of the Rydberg-Rydberg interaction, which impacts the population of highly excited Rydberg state giving rise to decoherence. Thus, selecting Rydberg states with high principle quantum number can help to block double excitation (green, dash lines) and reduce dissipation induced by spontaneous emission (blue, dash lines). Similar behavior happens for the case in which the system is initially in the state $|1\rangle_{1}|r\rangle_{2}$. 
 
Second, while the atoms 1 and 2 are in the states $|0\rangle_{1}|0\rangle_{2}$ and $|1\rangle_{1}|0\rangle_{2}$, the Rydberg blockade does not exist and conditional state transfer takes place. The laser pulses will lead to a resonant two-photon Raman transition for qubit 1. The time evolution can be found by setting $\Delta_{r}=0$ in Eq.~(\ref{eq:q12_CSTransfer}) with the initial state $|0\rangle_{1}|0\rangle_{2}$ or $|1\rangle_{1}|0\rangle_{2}$ and the related mediate system states. Choosing appropriate pulse length (see Fig.\ref{fig:Fig5}), we can realize the state transfer $|0\rangle_{1}|0\rangle_{2}\rightarrow|1\rangle_{1}|0\rangle_{2}$ and $|1\rangle_{1}|0\rangle_{2}\rightarrow|0\rangle_{1}|0\rangle_{2}$ perfectly without including spontaneous decay. The resonant transfer is fast compared with the dispersive interaction process. A practical spontaneous emission rate will slightly reduces the transferring efficiency. However, utilization of optimized pulse shapes will help to increase the fidelity. 

\begin{figure}
  \centering
\includegraphics[width=0.9\columnwidth]{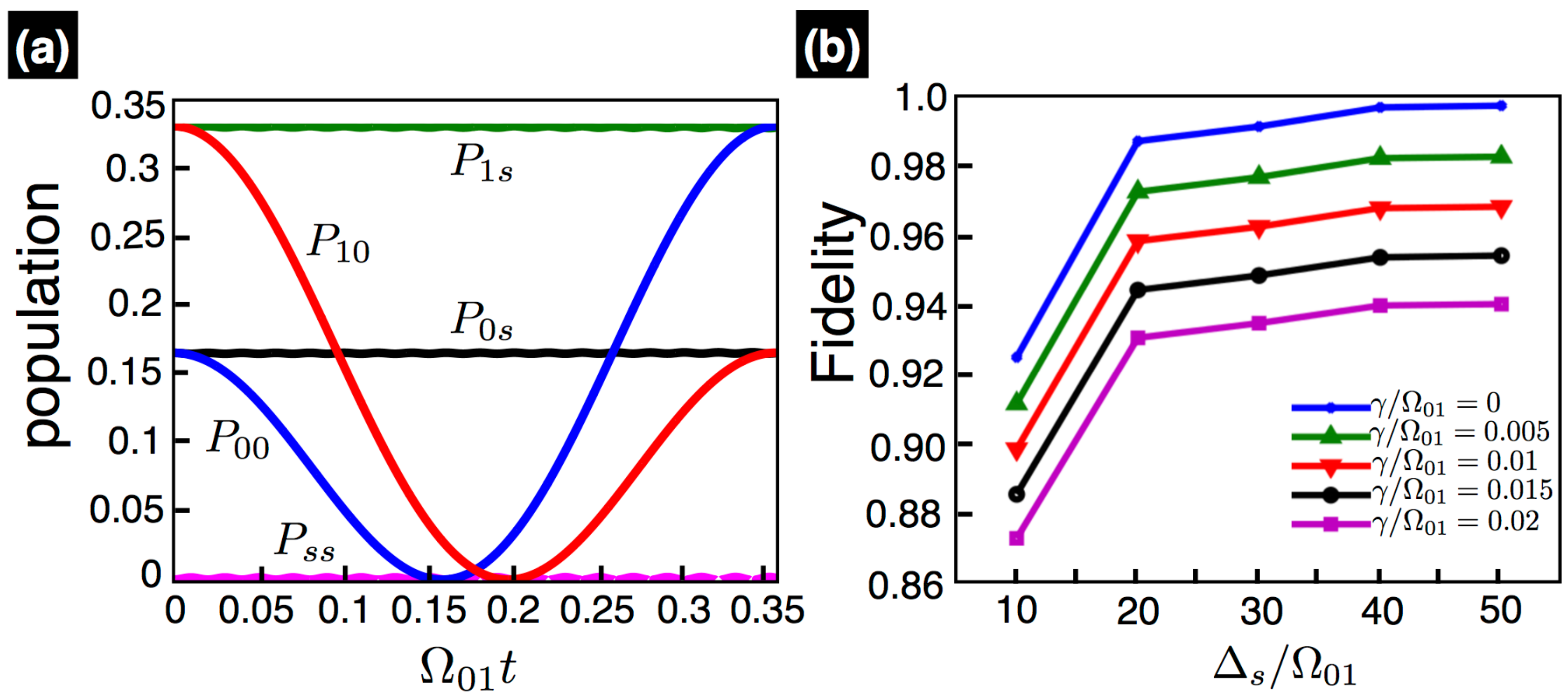}
\caption{\label{fig:Fig6} (a) Time-dependent population of the two-atom states $|00\rangle$, $|10\rangle$, $|0r\rangle$, $|1r\rangle$, and $|rr\rangle$ for $\Delta_r/\Omega_{01}=50$ and $\gamma=0$. (b) Fidelity of the CNOT-like transformation $M_{cnot}$ versus $\Delta_r/\Omega_{01}$ for varied spontaneous emission rate of the Rydberg state $|r\rangle$.}
\end{figure}

The laser pulse sequence actually implements a contolled NOT (CNOT) like quantum logic gate in only one step. The qubit 1 flips when qubit 2 is in the ground state $|0\rangle$. While qubit 2 is in the Rydberg state $|r\rangle$, the state transfer of the qubit 1 is blocked. The transformation is ideally given by 
 \begin{eqnarray}|0\rangle_{1}|r\rangle_{2}&\rightarrow&-|0\rangle_{1}|r\rangle_{2},\nonumber \\
 |1\rangle_{1}|r\rangle_{2}&\rightarrow&-|1\rangle|r\rangle_{2},\nonumber \\
 |0\rangle_{1}|0\rangle_{2}&\rightarrow&|1\rangle_{1}|0\rangle_{2},\nonumber \\
 |1\rangle_{1}|0\rangle_{2}&\rightarrow&|0\rangle_{1}|0\rangle_{2},\label{eq:cnot_transformation}\end{eqnarray}
 which can be described by an amplitude matrix: \begin{equation}M_{CNOT}=\left[\begin{array}{cccc}
-1 & 0 & 0 & 0\\
0 & -1 & 0 & 0\\
0 & 0 & 0 & 1\\
0 & 0 & 1 & 0
\end{array}\right]. \end{equation}
 It should be noted that the transformation can easily convert to CNOT gate operation with the help of the single qubit operation $|r\rangle_{2}\rightarrow-|r\rangle_{2}$. In fact, due to the finite Rydberg-Rydberg interaction strength, the state transfer can not be completely inhibited. Thus, the gate operation $M_{CNOT}$ is imperfect. As an example, we numerically calculate the gate fidelity by $F=Tr(\rho_{ideal}\rho_{12}(t))$ with the initial two-atom state $(|0\rangle_{1}+\sqrt{2}|1\rangle_{1})/\sqrt{3}\otimes(|0\rangle_{2}+|r\rangle_{2})/\sqrt{2}$, where $\rho_{ideal}$ is the density operator of the system after an ideally perfect gate operation $M_{CNOT}$ and $\rho_{12}$ is the density operator after the realistic  transformation including imperfections. The time-dependent density operator $\rho_{12}(t)$ is obtained by numerically simulating the master equation with the form $d \rho_{12}/dt=(1/i\hbar)[H_{12},\rho_{12}]+(\gamma/2)\sum_{j=1,2}\sum_{k=0,1}(2S^{(k)}_{j}\rho_{12}S_{j}^{(k) \dagger}-\rho_{12}S_{j}^{(k) \dagger}S^{(k)}_{j}-S_{j}^{(k) \dagger}S^{(k)}_{j}\rho_{12})$, here $S^{(k)}_{j}=|k\rangle_{jj}\langle r|$. As shown in Fig.\ref{fig:Fig6}(a), excluding the influence of dissipation, the population transfer of the two-atom states $|0\rangle_{1}|0\rangle_{2}$ and $|1\rangle_{1}|0\rangle_{2}$ are nearly perfect for $\Delta_{r}/\Omega_{01}=50$. The double excitation state $|r\rangle_{1}|r\rangle_{2}$ is successfully suppressed. We show the fidelity of the transformation as a function of blockade strength for a set of spontaneous emission rate in Fig.\ref{fig:Fig6}(b). For the Rydberg states of the lifetime $\tau>100\mu$s and interaction strength $\Delta_{r}>50$MHz, the gate fidelity can reach $0.97$ with the atom-laser coupling strength $\Omega_{01}$ around 1MHz. Thus, implementation of an accurate quantum computation is still in reach  with the gate error less than 0.03 \cite{Knill_Nature2005_3percent}.

 After the transformation, the Bell basis states referred to as maximally entangled states are disentangled, \begin{equation} |\Phi_{\pm}^{1,2}\rangle\rightarrow\frac{1}{\sqrt{2}}|1\rangle_{1}(|0\rangle_{2}\mp|r\rangle_{2}), \end{equation} 
\begin{equation}  |\Psi_{\pm}^{1,2}\rangle\rightarrow\frac{1}{\sqrt{2}}|0\rangle_{1}(|r\rangle_{2}\mp|0\rangle_{2}). \end{equation} 
 A single qubit operation with a $\pi/2$ pulse can transform $(|0\rangle_{2}+|r\rangle_{2})/\sqrt{2}$ and $(|0\rangle_{2}-|r\rangle_{2})/\sqrt{2}$ to $|0\rangle_{2}$ and $|r\rangle_{2}$, respectively \cite{Johnson_PRL2008}. Thus, the Bell states finally become

 \begin{equation}
|\Phi_{\pm}^{1,2}\rangle\rightarrow \left\{ \begin{array}{rcl}
|1\rangle_{1}|r\rangle_{2}\\
|1\rangle_{1}|0\rangle_{2}
\end{array}\right.,
\end{equation}

 \begin{equation}
 |\Psi_{\pm}^{1,2}\rangle\rightarrow \left\{ \begin{array}{rcl}
-|0\rangle_{1}|r\rangle_{2}\\
|0\rangle_{1}|0\rangle_{2}
\end{array}\right..
\end{equation} The joint measurement can be achieved by detecting qubits 1 and 2 separately. The lossless state-selective detection of individual neutral atom qubits trapped in optical lattice \cite{Gibbons_PRL2011_state_detection} or optical tweezer \cite{Fuhrmanek_PRL2011_state_detection} can be realized with high accuracy by probing laser induced fluorescence. To avoid the influence of spontaneous decay from Rydberg state, the qubit 2 can be firstly transferred to the ground state $|1'\rangle$. Then, the target atomic qubits can be addressed by using an off-resonantly polarized laser beam focused onto the individual atoms. The laser beam induces differential energy shifts for the relevant hyperfine ground states and tunes the target atoms into resonant with an external microwave field \cite{Weitenberg_Nature2011_singleSpinAdress}. The transfer between hyperfine ground states can be realized via the Landau-Zener transition leaving the other atomic qubits unaffected. The state-selective detection is finally applied to the target atoms.

(3) Single qubit operation. The outcomes of the measurement for qubit 1 and 2 is transmitted to the receiver. A corresponding local rotation $\sigma_{x}$, $\sigma_{y}$, $I$, $\sigma_{z}$ for the measurement outcomes $|1\rangle_{1}|r\rangle_{2}$, $|1\rangle_{1}|0\rangle_{2}$, $|0\rangle_{1}|r\rangle_{2}$, $|0\rangle_{1}|0\rangle_{2}$ respectively, is then made on the qubit 3 to reconstruct the initial state of qubit 1.

The teleportation scheme can be implemented with Rubidium atoms. Using polarized laser beams, we can couple ground states in $5s_{1/2}$ to an $ns$ or $nd$ Rydberg excited state mediated by $5p_{1/2}$ state based on the transition selection rule. Alternatively, we can in principle manipulate the atoms in the $5s$ ground states by illuminating them with a UV pulse resonant with a transition to an $np$ Rydberg state \cite{Farooqi_prl03_UV}. High-fidelity quantum teleportation needs strong Rydberg-Rydberg interaction, which arises from the large dipole moments of Rydberg atoms. The strength of the interaction can be tuned by using external electric fields \cite{Jaksch_PRL2000}. The extreme sensitivity to electric field makes it possible to control the Rydberg-Rydberg interaction via a mechanism referred as F\"orster interaction \cite{Forster_Physik1948}, which can be realized in the absence of applied electric fields. The interaction energy for two Rydberg-excited $^{87}$Rb atoms (with the principal quantum number $n=58$) separated by $4\mu$m can reach $\Delta_{r}/2\pi=50$MHz \cite{Wilk_PRL2010}. It is worthwhile to note that the interaction strength does not need to be precisely controlled for the implementation of the teleportation scheme. Moreover, the fidelity of the scheme is limited by the radiative lifetime of the Rydberg state. Spontaneous emission rate is determined by the principle and azimuthal quantum number of the Rydberg energy level and the temperature of the surrounding \cite{Saffman_RMP2010}. For $n>65$, the lifetimes of the $s$, $p$, $d$, and $f$ states are greater than $100\mu$s at room temperature \cite{Saffman_PRA2005}.

To estimate the total fidelity of the teleportation process, we have $|r=58d_{3/2}\rangle$ excited by a two-photon transition process and set the strengths for the coupling of the ground states to the Rydberg state $\Omega_{0r}=\Omega_{01}=\Omega=2\pi\times2.5$MHz \cite{Wilk_PRL2010}, which implies the conditions $\Delta_{r}/\Omega=20$ and $\gamma/\Omega\sim10^{-3}$. Then, the time needed for the preparation of the EPR state, the implementation of CNOT gate followed by a $\pi/2$ pulse, and the recovery of the original state on qubit 3 (a statistical average over the four different cases) are $t_{1}=\pi/2\sqrt{2}\Omega$, $t_{2}=\sqrt{2}\pi/\Omega+\pi/4\Omega$, and $t_{3}=\pi/\Omega$, respectively. Thus, the total time consumption of the scheme is about $T\approx600ns$. Based on the above parameters, we also find that the fidelity of the teleportation process can reach $0.976$ by assuming a fast and exact state selective measurement.

So far, we only consider the intrinsic errors, that is, the finite blockade shift and finite Rydberg lifetime, neglecting all other errors due to technical imperfection, such as the errors induced by spontaneous emission from the intermediate state $5p_{1/2}$ (with the rate $\Gamma=2\pi\times3$MHz) and the atomic motion. Now we assume that the two laser beams used for Rydberg excitation (two-photon resonance) are detuned by $\delta/2\pi=1$GHz from the intermediate level $5p_{1/2}$ and have the identical coupling strengths ($\sim 2\pi\times50$MHz) to the atoms, then the fidelity of the teleportation scheme reduces to 0.92 by including the spontaneous decay from the $5p_{1/2}$ state and assuming frozen motion of the atoms during the laser sequence \cite{Zhang_PRA12_FideRydGate,Gaetan_NP09_2RydCol}. Although we assume that the atoms are in motional ground state, however, the sub-Doppler temperatures at the level of $50\mu$K are sufficient for high-fidelity preparation of the EPR states and implementation of the quantum gates \cite{Saffman_PRA2005}. We have numerically checked that the fidelity of the teleportation process is only slightly modified by introducing a deviation induced by atomic motion to the desired condition $\Delta_r/\Omega\sim (1\pm0.1)\times20$. A relevant problem was recently discussed for two Rydberg-blockaded atom clouds by M\"obius et al. \cite{Mobius_PRA2013}. Another important problem is the Rydberg photoionization induced by trapping light, which presents a limit to the usable Rydberg lifetime and thus further decreases the fidelity of the scheme \cite{Saffman_PRA2005}.

Quantum teleportation can be used for construction of quantum gates \cite{Gottesman_Nature1999}. Making use of single qubit operations, Bell states measurements and Greenberger-Horne-Zeilinger (GHZ) states, the teleportation-based method is sufficient to built a universal quantum computer. The protocol proposed in Ref. \cite{Gottesman_Nature1999} has proved that the Hadamard gate which commuted through a Pauli gate produced a Pauli gate and similar property happens for CNOT gate and Toffoli gate. Thus, the nice merit of the teleportation construction is that the quantum gates only operate on specific known states, instead of operating on unknown states, before the Bell-basis measurements. A modified correction is made with single qubit operations at the end of the protocol to achieve preset quantum operation. Since the fidelity of the quantum gates can be tested before they actually work on the unknown quantum states, the failed operations can then be discarded intentionally \cite{Gottesman_Nature1999}. Although building a fault-tolerant quantum computer is still out of reach in the near future, performing quantum simulation with many-body system is definitely of experimental interest \cite{Bloch_NatPhy2012}. Teleportation based simulation method can be utilized for simulation of an interacting high-dimensional quantum system and precisely generation of many-body interaction terms \cite{Dur_PRA2008}. Quantum simulation can be regarded as an intermediate step to full scale quantum computer. Quantum computation or quantum simulation using teleportation can relax the experimental qualifications by reducing resource requirement \cite{Gottesman_Nature1999}.

 \begin{figure}
 \centering
\includegraphics[width=0.7\columnwidth]{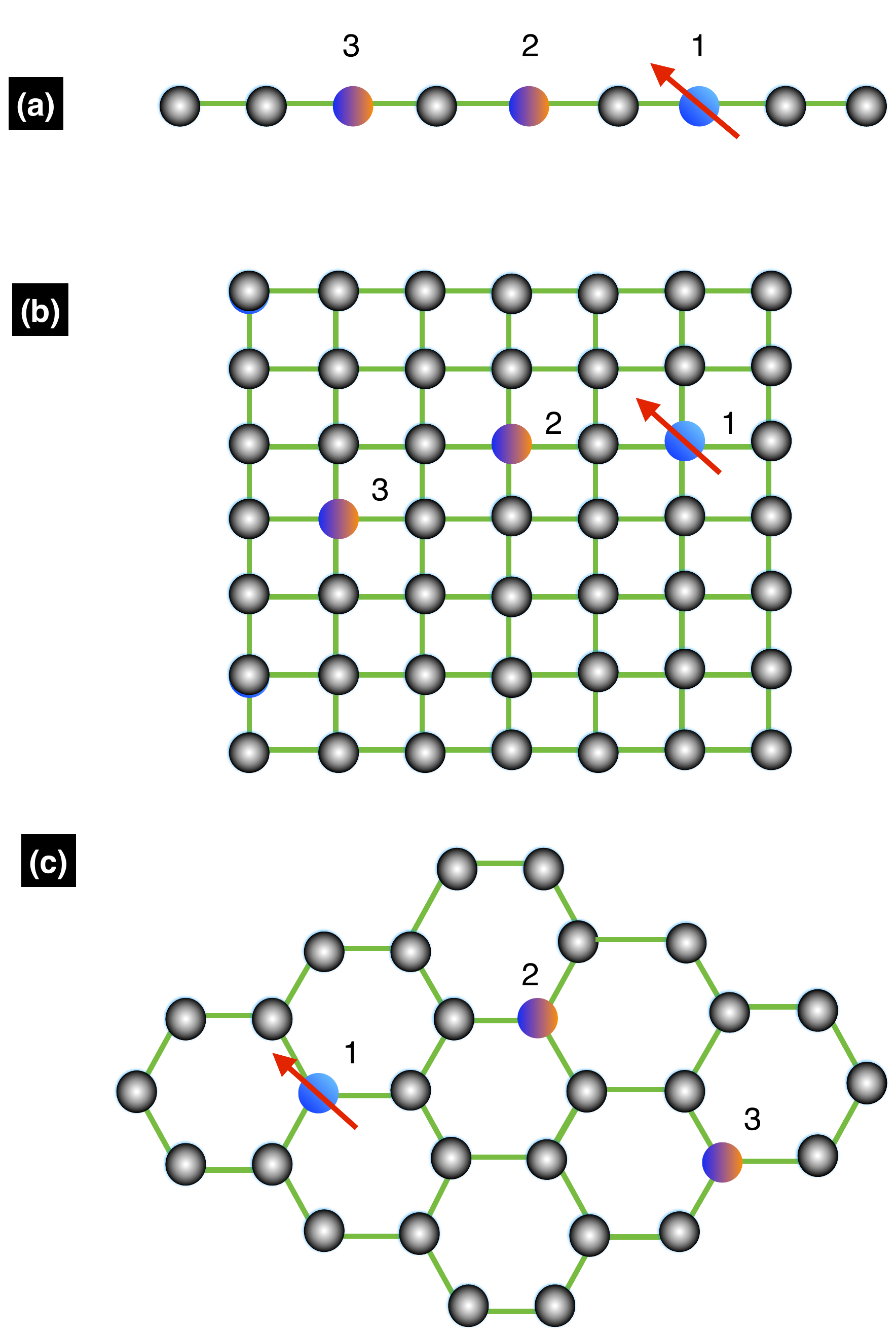}
\caption{\label{fig:Fig7} Trapped cold atoms in one- or two-dimensional optical lattice with different structures. Based on the Rydberg blockade mechanism, together with progressively experimental techniques in local operation and measurement, quantum teleportation can be implemented among the nodes of the whole atom network.}
\end{figure}

Rydberg atoms trapped in one- or two-dimensional optical lattice (Fig. \ref{fig:Fig7}(a) and (b)) \cite{Anderson_PRL2011} provides the possibility for doing quantum simulation \cite{Weimer_NatPhy2010} and adiabatic quantum computation \cite{Keating_arxiv2012}. Cubic lattices generated by superimposing three independent standing waves is the main way for trapping atoms. Generalized lattice structures including triangular and hexagonal \cite{Becker_njp2010}, which alternatively facilitate versatile control of Rydberg-Rydberg interaction. The simultaneous, state-insensitive confinement of the ground and Rydberg atoms in optical lattice can be realized by using the near-resonant blue-detuned trap \cite{Zhang_PRA11_MagicTrap,Li_Nat13_EnLightRydAtom}. For the purpose, the laser beams with the wavelength $1.012\mu$m are used for constructing the optical lattice, and the atoms are driven in resonance between the ground atomic level $5s_{1/2}$ and the Rydberg level $90s_{1/2}$ \cite{Li_Nat13_EnLightRydAtom}. In this case, the Rydberg atom size $\sim0.4\mu$m is smaller than the lattice period, the interactions between the highly excited electron of a Rydberg atom and a neighboring ground-state atom can be suppressed. The Rydberg-Rydberg interaction strength can reach several  hundred megahertz with the interatomic distance around $4\mu$m for the Rydberg state $90s_{1/2}$. As shown in Fig. \ref{fig:Fig7}, the atom to be teleported (in blue) is able to interact with the atoms (in optical lattice) within the Rydberg interaction range, according to which an interaction radius $r_{0}$ can be defined. Beyond $r_{0}$, the interatomic interaction is neglected. While the single-lattice-site resolved addressing is experimentally achieved, the Rydberg-Rydberg interaction can be selectively controlled between atom pairs in optical lattice. Suppose the distance between atoms 1 and 2 (or atoms 2 and 3) is less than $r_{0}$, while the interatomic distance between atoms 1 and 3 exceeds $r_{0}$. The excitation of atom 1 will only block atom 2 from excitation leaving atom 3 unaffected. In this case, the teleportation scheme can be easily realized. In the case where all the qubits involved in the teleportation scheme are within the interaction range $r_{0}$, then it is convenient to first transfer the atom 3 from the Rydberg state $|r\rangle$ to the ground state $|1\rangle$ before the Bell state measurement (i.e. the step (2)). This will be useful to avoid the many-body interaction and to reduce the influence of the finite radiative lifetime of the Rydberg state. Using the above rule, we can implement our teleportation protocol among all the lattice sites. A generalization of quantum teleportation can be then used for realization of fault-tolerant quantum computation \cite{Gottesman_Nature1999}. Compared to the quantum computing scheme where information is processed via a series of unitary gate operations, the teleportation-based method connecting to the measurement-based quantum computing scheme possesses the advantages that the quantum information can be processed via a sequence of adaptive measurement on an initially prepared, highly entangled resources state, and therefore, the resource requirements of the quantum information processing can be reduced \cite{Bloch_nature08_review}.

Recent progress in neutral atom experiments promises the successful implementation of the teleportation-based protocol. Single-site-resolved addressing and control of the individual atoms in a Mott insulator in an optical lattice has been demonstrated \cite{Weitenberg_Nature2011_singleSpinAdress}. The ground states of neutral atoms can couple to Rydberg excited states coherently \cite{Johnson_PRL2008}, and quantum entangled states and controlled NOT gate for two individual neutral atoms have been realized by using the Rydberg blockade method \cite{Isenhower_PRL2010,Wilk_PRL2010}. Preparation of GHZ states and implementation of multiqubit quantum phase gates based on the current architecture are theoretically feasible \cite{Saffman_PRL2009,Wu_PRA2010}. Therefore, useful transformations, including the set of gates in Clifford group, can be performed with quantum teleportation in a fault-tolerant way \cite{Gottesman_Nature1999}, which leads to conceptual simplification for a universal fault-tolerant quantum computer. A Rydberg atom array referred as Rydberg quantum simulator can reproduce the dynamics of the other many-body quantum systems \cite{Weimer_NatPhy2010}. Using entangled states as resources, teleportation-based gates are able to simulate the time evolution of the many-body Hamiltonian of the form $H=\sigma_{z}^{\otimes n}$ with $\sigma_{z}$ the Pauli matrix \cite{Dur_PRA2008}. 

In conclusion, we have studied the implementation of quantum teleportation with Rydberg neutral atoms. The Rydberg-excited atoms interact via the Rydberg blockade mechanism. The Rydberg-Rydberg interaction induced double-excitation energy shift and conditional state transfer are used for preparation of EPR pair and disentanglement, respectively. Using trapped neutral atoms (in optical lattice) as an architecture, quantum information can be teleported among the nodes of a quantum network. The teleportation-based quantum computation and quantum simulation (of interacting many-body system) can be carried out using quantum entangled states as resources. The protocol is based on the current experimental techniques and provides a new way for testing fundamental protocols in quantum information science and studying basic phenomena in condense matter physics. 

{\it Acknowledgements.} - This work was supported by the Major State Basic Research Development Program of China under Grant No. 2012CB921601, the National Natural Science Foundation of China under Grant No. 11247283, the Natural Science Foundation of Fujian Province under Grant No. 2013J01012, and the fund from Fuzhou University.   
\section*{References}

\bibliographystyle{jphysicsB}
%\bibliographystyle{unsrt}
%\begin{thebibliography}{<46>}
\bibliography{/Users/huaizhiwu/Lab_Huaizhi/Research_Projects/RydbergQC_Teleportation/Ryd_Tel_citations}
%\end{thebibliography}
\end{document}